\documentstyle[12pt]{article}

\def\mgravitino{m_{3/2}}

\catcode`\@=11
\long\def\@makefntext#1{
\protect\noindent \hbox to 3.2pt {\hskip-.9pt
$^{{\ninerm\@thefnmark}}$\hfil}#1\hfill}		

\def\@makefnmark{\hbox to 0pt{$^{\@thefnmark}$\hss}}  
\def\ps@myheadings{\let\@mkboth\@gobbletwo
\def\@oddhead{\hbox{}
\rightmark\hfil\ninerm\thepage}
\def\@oddfoot{}\def\@evenhead{\ninerm\thepage\hfil
\leftmark\hbox{}}\def\@evenfoot{}
\def\sectionmark##1{}\def\subsectionmark##1{}}

\renewcommand{\thefootnote}{\fnsymbol{footnote}}

\newcounter{sectionc}\newcounter{subsectionc}\newcounter{subsubsectionc}
\renewcommand{\section}[1] {\vspace*{0.6cm}\addtocounter{sectionc}{1}
\setcounter{subsectionc}{0}\setcounter{subsubsectionc}{0}\noindent
	{\normalsize\bf\thesectionc. #1}\par\vspace*{0.4cm}}
\renewcommand{\subsection}[1] {\vspace*{0.6cm}\addtocounter{subsectionc}{1}
	\setcounter{subsubsectionc}{0}\noindent
	{\normalsize\it\thesectionc.\thesubsectionc. #1}\par\vspace*{0.4cm}}
\renewcommand{\subsubsection}[1]
{\vspace*{0.6cm}\addtocounter{subsubsectionc}{1}
	\noindent {\normalsize\rm\thesectionc.\thesubsectionc.\thesubsubsectionc.
	#1}\par\vspace*{0.4cm}}

\newcounter{appendixc}
\newcounter{subappendixc}[appendixc]
\newcounter{subsubappendixc}[subappendixc]

\renewcommand{\appendix}[1] {\vspace*{0.6cm}
        \refstepcounter{appendixc}
        \setcounter{figure}{0}
        \setcounter{table}{0}
        \setcounter{equation}{0}
        \renewcommand{\thefigure}{\Alph{appendixc}.\arabic{figure}}
        \renewcommand{\thetable}{\Alph{appendixc}.\arabic{table}}
        \renewcommand{\theappendixc}{\Alph{appendixc}}
        \renewcommand{\theequation}{\Alph{appendixc}.\arabic{equation}}
        \noindent{\bf Appendix \theappendixc #1}\par\vspace*{0.4cm}}

\def\abstracts#1{{
\centering{\begin{minipage}{12.2truecm}\vspace*{.1cm}
        \footnotesize\baselineskip=12pt\noindent
	\parindent=0pt #1
	\end{minipage}}\par}}

\renewenvironment{thebibliography}[1]
	{\begin{list}{\arabic{enumi}.}
	{\usecounter{enumi}\setlength{\parsep}{0pt}
\setlength{\leftmargin 1.25cm}{\rightmargin 0pt}
	 \setlength{\itemsep}{0pt} \settowidth
	{\labelwidth}{#1.}\sloppy}}{\end{list}}

\newcounter{itemlistc}
\newcounter{romanlistc}
\newcounter{alphlistc}
\newcounter{arabiclistc}

\newcommand{\fcaption}[1]{
        \refstepcounter{figure}
        \setbox\@tempboxa = \hbox{\footnotesize Fig.~\thefigure. #1}
        \ifdim \wd\@tempboxa > 6in
           {\begin{center}
        \parbox{6in}{\footnotesize\baselineskip=12pt Fig.~\thefigure. #1}
            \end{center}}
        \else
             {\begin{center}
             {\footnotesize Fig.~\thefigure. #1}
              \end{center}}
        \fi}

\newcommand{\tcaption}[1]{
        \refstepcounter{table}
        \setbox\@tempboxa = \hbox{\footnotesize Table~\thetable. #1}
        \ifdim \wd\@tempboxa > 6in
           {\begin{center}
        \parbox{6in}{\footnotesize\baselineskip=12pt Table~\thetable. #1}
            \end{center}}
        \else
             {\begin{center}
             {\footnotesize Table~\thetable. #1}
              \end{center}}
        \fi}

\def\@citex[#1]#2{\if@filesw\immediate\write\@auxout
	{\string\citation{#2}}\fi
\def\@citea{}\@cite{\@for\@citeb:=#2\do
	{\@citea\def\@citea{,}\@ifundefined
	{b@\@citeb}{{\bf ?}\@warning
	{Citation `\@citeb' on page \thepage \space undefined}}
	{\csname b@\@citeb\endcsname}}}{#1}}

\newif\if@cghi
\def\cite{\@cghitrue\@ifnextchar [{\@tempswatrue
	\@citex}{\@tempswafalse\@citex[]}}
\def\citelow{\@cghifalse\@ifnextchar [{\@tempswatrue
	\@citex}{\@tempswafalse\@citex[]}}
\def\@cite#1#2{{$\null^{#1}$\if@tempswa\typeout
	{IJCGA warning: optional citation argument
	ignored: `#2'} \fi}}

 1
 1
 1

\font\ninerm=cmr9

\begin{document}

\rightline{SLAC-PUB-95-6934}
\vskip.4in

\centerline{\normalsize\bf SUPERSYMMETRY IN THE VERY EARLY
UNIVERSE\footnote{To appear in the proceedings of PASCOS/HOPKINS 1995.}
}
\baselineskip=22pt
\centerline{\footnotesize SCOTT THOMAS\footnote{Work Supported
by the Department of Energy under contract DE-AC03-76SF00515.}
}
\baselineskip=13pt
\centerline{\footnotesize\it Stanford Linear Accelerator Center}
\baselineskip=12pt
\centerline{\footnotesize\it Stanford University}
\baselineskip=12pt
\centerline{\footnotesize\it Stanford, CA  94309}
\vspace*{0.5cm}

\vspace*{0.9cm}
\abstracts{
Supersymmetric flat directions can have a number of important
consequences in the very early universe.
Depending on the form of the SUSY breaking potential
arising from the finite energy density
at early times,
coherent production of scalar condensates can result along
such directions.
This leads a cosmological disaster for Polonyi type flat directions
with only Planck suppressed couplings, but can give rise to the
baryon asymmetry for standard model flat directions.
Flat directions are also natural candidates to act as inflatons.
Achieving density fluctuations of the correct magnitude
generally requires
an additional hidden SUSY breaking sector.
}

\normalsize\baselineskip=15pt
\setcounter{footnote}{0}
\renewcommand{\thefootnote}{\alph{footnote}}

\section{Introduction}

At present,
supersymmetry seems to provide the most likely solution to the
problem of the large hierarchy between the weak and GUT or Planck
scales.
As there is yet no direct experimental evidence for supersymmetry,
it is worth turning to the early universe for possible signatures.
In this review I will describe some recent progress in understanding
effects of supersymmetry in the {\it very} early
universe.
Here the relevant measure for the epoch of the early universe
will be the Hubble constant.
Most of the important effects discussed below
will be for $H > 100$ GeV.
(This is to be compared with, for example, the electroweak phase
transition which takes place 
when $H \sim 10^{-12}$ GeV).
At such high energy scales we almost certainly don't
know the full spectrum or interactions.
In order to glean any information there must
be some affects associated with
a generic feature of supersymmetric theories, which is
not shared by non-supersymmetric theories.
Flat directions are just such a general feature of supersymmetric
theories.
As described in more detail below, flat directions can
lead to the coherent production of scalar condensates.
These condensates can have important consequences, including
the production of the baryon asymmetry.
In addition, flat directions are natural candidates to
act as inflatons.
This generally requires an additional sector which breaks supersymmetry
at a high scale.
These considerations may even give some hints to possible
cosmological selection principles for the type of vacuum
in which we live.


Flat directions are directions in field space on which the perturbative
potential exactly vanishes.
Such directions arise essentially as accidental classical degeneracies.
Supersymmetric theories are special in that the non-renormalization
theorem guarantees that these degeneracies are not lifted at any
order in perturbation theory.
Without supersymmetry in general an accidental degeneracy would be
lifted quantum mechanically.
Flat directions are quite common in supersymmetric theories.
For example, in the minimal supersymmetric standard model there is a
37 complex dimensional subspace of the full field space on which
the scalar gauge potential vanishes.
In this subspace there are 100's of rays on which the renormalizable
potential arising from Yukawa couplings vanishes.
In string theory flat directions are also common.
The internal manifold on which the two dimensional fields live
often
depends on a continuous set of parameters, ${\cal M}(\phi)$, which
leave the theory conformally invariant.
Geometrically the $\phi$ may be thought of as
deformations of ${\cal M}$.
The operators which describe the deformations are exactly marginal,
and so in four dimensions
appear as exactly flat directions.
The space of all flat directions in a theory is usually referred
to as the moduli space.

Fields coupled by renormalizable interactions
gain a large mass along flat directions.
The moduli space therefore contains the relevant degrees of
freedom to describe the evolution of fields in the early universe.

\section{Supersymmetry Breaking in the Early Universe}

Important for all the cosmological effects discussed below is
the potential along flat directions.
As it is supersymmetry which protects an exact flat direction from
obtaining a potential, in the presence of SUSY breaking a potential
results.
Throughout I will make a hidden sector assumption in which
the intrinsic SUSY breaking is transmitted to flat directions
only through gravitational strength interactions.
This turns out to be justified far out along flat directions
even if there are other interactions (except in special cases).
The general form of the soft SUSY breaking potential is then
\begin{equation}
V(\phi) = m^2 M_p^2 {\cal F}(\phi/M_p)
\label{flatpot}
\end{equation}
where $\phi$ parameterizes the flat direction,
$M_p$ is the Planck mass,
$m \sim \Lambda^2/ M_p$ is the soft SUSY breaking mass, and
$\Lambda$ is the intrinsic SUSY breaking scale.
Under the assumptions given above, in the present universe
$\Lambda_{\rm now} \sim 10^{11}$ GeV in a hidden SUSY breaking
sector, giving $m \sim \mgravitino \sim 10^2 - 10^3$ GeV.
However in the early universe SUSY breaking
can arise from other sources.
In particular the finite energy density, $\rho$, necessarily generates
a SUSY breaking potential along flat directions with
$\Lambda_{\rho}^2 \sim \sqrt{\rho}$.\cite{drt}
Using the relation between $H$ and $\rho$ for an expanding background,
$\rho = 3 H^2 M_p^2$, implies that $m \sim H$ for $H$ above the
weak scale.\cite{drt,dvali}
In general, the specific functional form of the potential from
this source does not coincide with that from hidden sector SUSY breaking.
This has important implications for the coherent production of
condensates, as described in the next section.
Also, the existence of another hidden sector with
$\Lambda \sim 10^{16}$ GeV can lead to successful inflation along
flat directions as discussed in section 4.

\section{Coherent Production of Scalar Fields}

The evolution of a flat direction in the early universe
is determined by the classical equations of motion.
The equation of motion for the average value of $\phi$ is
\begin{equation}
\ddot\phi + 3 H \dot\phi + V^{\prime}(\phi) =0
\end{equation}
The damping term, proportional to $H$, arises because of the expanding
background.
For $H^2 \gg V^{\prime \prime}$ the field is overdamped,
while for $H^2 \ll V^{\prime \prime}$ it is underdamped.
Previously, it had been implicitly assumed that the potential along a
flat direction was set by hidden sector SUSY breaking, i.e.
$V^{\prime \prime} \sim \mgravitino^2 \sim (100$ GeV)$^2$.
If this were the case, then at early times the field would be highly
overdamped and effectively frozen.
Since the hidden sector potential is much smaller than the relevant
mass scale at very early times, it would seem reasonable to assume
that the initial value of the fields along a flat direction is
random.
When the Hubble constant decreases
to $H \sim \mgravitino$, the field
would begin to oscillate freely about a minimum of the potential.
These oscillations redshift like matter and amount to a coherent
condensate of nonrelativistic particles.
Since flat directions are a generic feature of supersymmetric theories,
the production of scalar condensates in the early universe would
also seem to be a generic feature.
A number of questions about this scenario immediately arise
however.
Among these are the questions of initial conditions and
the form of the potential along flat directions at early
times.

The SUSY breaking potential arising from the finite energy
density has an important effect on the production of condensates,
and helps to answer the questions raised above.
Since $m \sim H$ at early times,
rather than being highly overdamped, the field is always
parametrically near critically damped.
During inflation when $H$ is roughly
constant, the field is then driven very efficiently
to an instantaneous minimum of the potential.
However, the typical scale of variation in the soft potential
is $M_p$.
So the minimum of the finite density induced potential
is in general displaced by ${\cal O}(M_p)$ from the
true minimum for a direction which is exactly flat in the SUSY limit.
For a direction which is lifted by non-renormalizable terms in the
superpotential, $\phi \ll M_p$ just on energetic grounds, but
large displacements can still occur.
In either case,
when $H\sim \mgravitino$ the field then begins to oscillate
in the true potential
with a large ``initial'' expectation value and a condensate is formed.
The subsequent evolution of the condensate depends on its quantum numbers
and couplings.
Two examples are the production of a condensate of Polonyi type
moduli, and the generation of a baryon asymmetry along
flat directions of standard model fields.





\subsection{The Polonyi Problem}

Polonyi fields
are flat directions which are exactly flat in the supersymmetric limit
and
have only gravitational strength
couplings to light fields.
Fields of this type are common in hidden sector models of SUSY breaking,
and the moduli of string theory fall in this class.
Since these directions are exactly flat in the SUSY limit,
as discussed above,
the finite density induced potential in general causes the fields
to be displaced by ${\cal O}(M_p)$ from the true minimum.
The resulting condensate then dominates the energy density
essentially as soon as free oscillations begin when $H \sim \mgravitino$.
Since the condensate has only Planck suppressed interactions
its lifetime is quite long, $\tau \sim 8 \pi^2 M_p^2/m^3 \sim 10^4$ s.
This leads to a number of cosmological problems.
The decay takes place during and after the era of nucleosynthesis.
Photodissociation and photoproduction modifies the light element
abundances in an unacceptable manner.
In addition, if the LSP is stable, the relic density is far too large.
Finally, because of the large entropy release,
baryogenesis must take place during or after the decay.
The production of Polonyi condensates is obviously a cosmological
problem which must be avoided in some way.

There have been a number of suggestions to solve the Polonyi problem.
If the minimum of the finite density induced potential happened
to coincide with the true minimum arising from hidden sector
SUSY breaking, the moduli would be driven to a stationary point
during inflation, and no condensate would result after
inflation.\cite{drt,dvali}
This is technically natural if there is a
point of enhanced symmetry
on moduli space.\cite{drt}
The potential is always extremum about such points and could be
a minimum at both early and late times if the enhanced symmetry
remains unbroken (aside from possible weak scale breaking now).
Such points are in fact common in string theory, and are
analogous to the self dual radius for toroidal compactification.
For the dilaton the only candidate for such a symmetry
seems to be $S$ duality.
However, at the dilaton self dual point
the four dimensional gauge coupling is likely to be very large.
So if symmetries are the solution to the Polonyi problem,
the dilaton is probably on a different footing.
Another possibility is that the dangerous Polonyi directions
gain a mass from SUSY preserving dynamics at a high scale,
and therefore decay at an early epoch.
This solution is natural in the context of moduli inflation
(described in section 4) which requires an additional dynamical
sector to drive inflation.\cite{bbs}
Again the dilaton is problematic within this solution.
Since SUSY breaking and the superpotential
in the sector responsible for driving
inflation must vanish at the minimum of the inflaton potential,
the dilaton potential at late times can not arise from this
sector.\cite{bbs}
The only know exceptions to this are racetrack schemes which
stabilize the dilaton through a balance between multiple
dynamical sectors.\cite{racetrack}
Another possibility is that a brief period of late inflation
sufficiently dilutes the Polonyi fields, while retaining
density fluctuations on large scales.\cite{weakinf}

Lyth and Stewart have recently suggested a version of late inflation
which has a number of desirable features.\cite{lsgut}
The main assumption is that a flat direction has a
potential (arising from hidden sector SUSY breaking) with a
minimum at some scale $M \ll M_p$, and that the
cosmological constant vanishes at this minimum.
The scale $M$ might be associated with a GUT or intermediate scale.
In addition the origin ($\phi =0$)
is assumed to be a maximum for this direction.
Now at high temperatures the minimum of the free energy can
reside at the origin.
This can occur if additional states become massless, thereby
increasing the entropy contribution.
So at early times it is possible for the flat direction to be
held at the origin by thermal effects.
{}From the form of the potential (\ref{flatpot}), and with
the assumptions spelled out above, there is a constant
contribution to the vacuum energy at this point on moduli space of
$V_0 \sim \mgravitino^2 M^2$.
Once the energy density of the Polonyi condensate
and thermal
plasma drop below this value, the universe enters a period
of inflation with Hubble constant $H \sim \mgravitino (M/M_p)$.
The temperature drops as the thermal plasma is diluted during
inflation.
When $T \sim \mgravitino$ the thermal effects no longer trap the
flat direction at the origin, and inflation ceases.
The flat direction then begins to oscillate about the true
minimum.
Theses oscillations eventually decay to standard model fields
through non-renormalizable operators suppressed by the scale $M$.
In order for the decays to take place before the era of
nucleosynthesis, $M$ probably can not be associated directly
with the GUT scale, but could be related to an intermediate
scale.\cite{bdt}
The total number of $e$-foldings during
inflation is
$N \sim \ln(T_i/\mgravitino)$ where $T_i$ is the temperature
when $V_0$ dominates the energy density,
$g_* T_i^3 T_R \sim \mgravitino^2 M^2$, and $T_R$ is the reheat
temperature from an earlier standard inflation.\cite{bdt}
Under reasonable assumptions, $N$ is large enough to sufficiently
dilute the Polonyi fields, but small enough to avoid
eliminating primordial density fluctuations on large
scales.
This type of inflation seems to be unique in that it necessarily
undergoes a small number of $e$-foldings,
and so is well suited to solve the Polonyi problem.
Since no fields undergo Planck scale excursions, and $H \ll \mgravitino$
during this inflation,
the form of the moduli potential essentially
coincides with the true potential, and the Polonyi fields are driven
to the true minimum.
Gravitinos are also diluted by this late inflation.
Because of the large dilution, baryogenesis probably has to take
place in the decay of the oscillating flat direction after inflation.
This last requirement probably requires the explicit or spontaneous
breaking of $R$ parity.\cite{bdt,latelsp}

\subsection{Baryogenesis Along Flat Directions}

Individual flat directions necessarily carry a global quantum
number.
For standard model flat directions the possibility therefore
exists to store a net baryon number in a condensate.
The production of a net asymmetry in a condensate depends on
the magnitude of the $B$ violating terms in the
potential when the
direction begins to oscillate freely ($H \sim \mgravitino$).
Affleck and Dine originally suggested this as a mechanism
of baryogenesis.\cite{ad}

Standard model directions which are flat at the
renormalizable level will in general be lifted by
non-renormalizable interactions in the superpotential,
\begin{equation}
W = {\lambda \over n M^{n-3}} \phi^n
\end{equation}
where $\phi$ parameterizes the direction made of $n$
standard model fields, and $M$ is some large mass scale such
as the GUT or Planck scale.
Given the discussion of finite density
SUSY breaking in section 2,
the relevant part of the potential along $\phi$ is
\begin{equation}
V(\phi) = (c H^2 + m_{\phi}^2) |\phi|^2
 + \left( {(A + a H) \lambda \phi^n \over n M^{n-3}} ~ + h.c. \right)
 + | \lambda|^2 {|\phi|^{2n-2} \over M^{2n-6} }
\end{equation}
where $m_{\phi} \sim A \sim \mgravitino$ are soft
parameters arising from hidden sector SUSY breaking, and
$c \sim a \sim {\cal O}(1)$ are the soft parameters induced
by the finite energy density.\cite{drtbaryo}
The $A$ term (proportional to $W$) has the important effect
of violating $B$ and has a definite $CP$ violating phase
relative to $\phi$.
Notice that the question of an enhanced symmetry point does not
arise for standard model flat directions.
The origin is always a symmetry point and the potential is an extremum
there.
If $c > 0$ the field gets driven to the origin during
inflation and no condensate forms.
However, if $c <0$ the minimum at early times lies at
$|\phi| \sim ( HM^{n-3} / \lambda)^{1/(n-2)} \ll M_p$.
During inflation the field gets driven very close to this value.
After inflation $H$ decreases and the instantaneous value of the minimum
decreases in time.
The field tracks close to this value during this epoch.
Eventually when $H \sim \mgravitino$ the $m_{\phi}^2$ term
from hidden sector SUSY breaking dominates, and the field begins
to oscillate about the new minimum at $\phi=0$.
However, just at this time the magnitude of the $A$ term is
necessarily the same order as the other terms in the potential.
A near maximal asymmetry (depending precisely on the initial
phase of $\phi$) therefore results in the condensate.\cite{drtbaryo}
Note that this is {\it independent} of the magnitude non-renormalizable
operator which lifts the flat direction, and any initial condition
assumptions.

Even though the fractional
asymmetry is largely independent of any details,
the total density in the condensate depends on the order at which
the flat direction is lifted.
In addition, the relevant quantity is the baryon per entropy
ratio.
Putting everything together, $n_b/s$ depends mainly on the reheat
temperature after inflation, $T_R$, and the order at which
the direction is lifted
\begin{equation}
{n_b \over s} \sim \epsilon ~{T_R \over m_{\phi}}
\left( { \mgravitino M^{n-3} \over \lambda M_p^{n-2}} \right)^{2/(n-2)}
\end{equation}
where $\epsilon \sim {\cal O}(1)$ is the condensate asymmetry, and
the quantity in brackets is the ratio $\rho_{\phi}/\rho_{tot}$
when $\phi$ begins to oscillate freely.
Unless $T_R$ is near the weak scale,
$n_b/s$ is too large for $n \geq 6$ without additional entropy releases.
However for $n=4$ a reasonable value results,
$n_b/s \sim 10^{-10}(T_R / 10^9~{\rm GeV})
(M / \lambda M_p)$.
In the MSSM with conserved $R$ parity, $LH_u$ is the unique
renormalizable flat direction
which is lifted at $n=4$ and which
carries $B-L$ (the non-anomalous combination of $B$ and $L$).
This direction is also special in that it is the only one which
has a $H_u$ component, and is therefore perhaps the most likely
to develop a negative mass squared in the early universe
from renormalization group evolution
(because of the large top quark Yukawa).
The operator which lifts this direction,
$W=(\lambda/M)(LH_u)^2$, is responsible at low energies for
neutrino masses.
So in this scenario the baryon asymmetry is related to lightest
neutrino mass,
$n_b/s \sim 10^{-10} (T_R/10^9~{\rm GeV})(10^{-5}~{\rm eV}/m_{\nu})$.



It is also possible to obtain a baryon asymmetry from non-standard
model flat directions.
In fact, if a condensate decays through non-renormalizable operators
which couple to $R$-odd combinations of standard model fields,
both a baryon asymmetry and relic LSPs can be produced.\cite{latelsp}
In this scenario the baryon and dark matter densities have
the natural relation
$\Omega_b / \Omega_{\rm LSP} \sim m_b / m_{\rm LSP}$.
Assuming closure density, limits on the baryon density from
nucleosynthesis then give an upper limit on the LSP mass
in this scenario,
$m_{\rm LSP} < 100h^2$ GeV, where
$h=H/(100~{\rm km}~s^{-1}~{\rm Mpc}^{-1})$.\cite{latelsp}

\section{Inflation Along Flat Directions}


The existence of an inflationary phase in the early universe
eliminates the flatness and horizon problems of standard
big bang cosmology.
In addition, quantum deSitter fluctuations of the inflaton field driving
inflation give rise to a (nearly) scale invariant spectrum of
density fluctuations, which can act as seeds for structure formation.
In order to be consistent with density and temperature
fluctuations in the present
universe, $\delta \rho / \rho \sim \delta T/T \sim 10^{-5}$,
the inflaton potential
must be extremely flat, with a very small dimensionless self
coupling, $\lambda \sim 10^{-8}$.
Since the potential for moduli are exactly flat in the supersymmetric
limit, these fields seem to be natural candidates to act as
inflatons.\cite{bg,moduliinfa,moduliinfb}
A nontrivial vacuum energy (as required for inflation)
over moduli space requires SUSY breaking.
The most straight forward way in which this can arise
is for there to be nontrivial moduli dependence
of a SUSY breaking scale, $\mu$, in some sector.
Assuming only Planck scale couplings between this SUSY breaking
sector and moduli, the form of the moduli potential is then
\begin{equation}
V(\phi) = \mu^4 {\cal F}(\phi/M_p)
\label{inflatonpot}
\end{equation}
The small self coupling arises from the ratio of mass scales,
$\lambda \sim (\mu / M_p)^4$.
The correct magnitude for density and temperature fluctuations results for
$\mu \sim 10^{16}$ GeV, giving a Hubble constant during inflation
of $H \sim 10^{14}$ GeV.
With the form of the potential (\ref{inflatonpot}) the power in
gravitational waves is naturally much smaller than that in
scalar perturbations.

It is often remarked that the small self coupling
of inflaton potentials amounts to extreme fine
tuning.
However, here $\mu$ arises from dynamical SUSY breaking as
the result of dimensional transmutation, and can be
hierarchically smaller than the Planck scale.
No fine tuning is required (aside from the moderate amount of
tuning to achieve slow roll so that the vacuum energy
dominates the kinetic energy during inflation).
It is even possible for the inflaton to have renormalizable
couplings to other fields.
The non-renormalization theorem guarantees that even in the
presence of such large couplings, the form of the potential
(\ref{inflatonpot}) is not modified.
Flat directions in standard model fields could even
act as inflatons.\cite{weakinf,moduliinfa}

In order for moduli to act as inflatons with a SUSY breaking
potential of the scale given above, a number of
interesting requirements must be met.
The first is that SUSY breaking in the sector responsible for
driving inflation should vanish at the minimum of the moduli
potential, i.e.
$DW(\phi_0)=0$ where W is the nonperturbative superpotential
generated over moduli space, and $D$ is the Kahler derivative
(the local version of the field derivative in global SUSY).
If this were not the case the large SUSY breaking would remain after
inflation, giving a gravitino mass (an therefore weak scale)
of $\mgravitino \sim \mu^2 / M_p \sim 10^{14}$ GeV.
The second requirement comes from the form of the supergravity
potential
\begin{equation}
V=e^K \left( DW \bar{D}W^* - {3 \over M_p^2} |W|^2 \right)
\end{equation}
Since $DW$ must vanish at the minimum, if
$W(\phi_0) \neq 0$ the cosmological constant is negative after
inflation.
As spelled out very clearly by Banks, Berkooz, and Steinhardt,
if this were the case the universe enters a phase of
irreversible contraction.\cite{bbs}
Therefore there must be a special point on moduli space
at which $DW=W=0$ in the sector responsible for driving inflation.
Alternately, the only type of vacua which exit inflation and
remain large have this property.\cite{bbs,moduliinfb}

Two possibilities for satisfying these requirements have
been suggested.
The first is to assume that the SUSY breaking responsible for driving
inflation arises from the nonperturbative modification of
a classical singularity for a composite field.\cite{moduliinfa}
If $X$ is a composite flat direction made of $n$ fields,
then its Kahler potential has a classical singularity at the origin,
$K=(X^{\dagger}X)^{1/n}$.
Under some circumstances it is believed that this singularity can
be smoothed out (if the composite field satisfies all the
t'Hooft anomaly matching conditions at the origin) giving
$K \simeq (X^{\dagger}X)/\Lambda^{2n-2}$ where $\Lambda$
is the dynamical scale.
In the presence of a superpotential, $W = \beta X/M_p^{n-3}$,
this leads to SUSY breaking with vacuum energy
$V=\beta^2 \Lambda^{2n-2} / M_p^{2n-6}$, with $X=0$.
At present there is only one know model of this type,
but there are probably others.\cite{ttmodel}
If the coefficient of the superpotential
is moduli dependent, $\beta=\beta(\phi)$, a potential
results over moduli space.
However, since $X=0$ is the stable point, $W=0$ over all
of moduli space.
The only requirement is then that $DW$ vanishes at some point,
which happens if $\beta(\phi)$ has a zero.
The second suggestion for satisfying the requirements is
to assume that a nonperturbative superpotential is generated
over moduli space $W=W(\phi)$,
and that both $W$ and $DW$ vanish at some
point.\cite{bbs} 
As an example, an $SU(N_c)$ gauge theory with
$N_f < N_c$ flavors of vector matter transforming in the
fundamental
gives rise to a nonperturbative superpotential.
If the masses of the $N_f$ flavors are moduli dependent,
a nontrivial moduli potential results.
If there is an enhanced symmetry point on moduli space, $\phi_0$, at
which additional flavors become massless such that $N_f > N_c$,
then $W(\phi_0)=DW(\phi_0)=0$ and the above requirements are satisfied
in a technically natural way.\cite{bbs}

As these two examples illustrate, it is possible for an additional
SUSY breaking sector (beyond the one required to give the
``observed'' splitting within the standard model multiplets)
to generate an acceptable inflaton potential.
It is interesting to note that with the above conditions, the
cosmological constant naturally vanishes after inflation;
no tuning of the overall zero of the potential is required to
exit inflation.
It is also natural within moduli inflation to implement
the suggestion of Linde and Vilenkin that topological defects
can act as seeds for inflation.\cite{seeds}
String moduli transform under modular symmetries and
can support topological defects.\cite{moduliinfb}
Finally, for composite flat directions which act as inflatons,
the reheat temperature after inflation can in principle
be much lower
than for standard singlet inflatons.\cite{moduliinfa}

The dilaton flat direction of string theories has always presented
a cosmological dilemma.
Since its expectation value is inversely proportional to the
gauge coupling constant, its potential goes to zero at large
value.
This leads to the runaway dilaton problem.\cite{bs}
Now there must be some barrier for the dilaton between
very weak coupling and its true minimum now.
Inflation from dynamical SUSY breaking then gives a partial
resolution of the problem.
In regions of the universe where the dilaton is on the
strong coupling side of the barrier, the potential during
inflation can in principle
keep the dilaton within the basin of attraction
of the true minimum.
In regions where the dilaton is on the weak coupling side of the
barrier it is driven to very weak coupling.
The dynamical sector which drives inflation then also gets driven to
weak coupling, and successful inflation never completes.
The only regions of the universe which undergo sustained inflation and
get big are on the strong coupling side.
It is also possible (on the strong coupling side)
for the dilaton itself to
be part of the inflaton direction.\cite{moduliinfb}








\section{Conclusions}

Supersymmetric flat directions
apparently play a important role in the very early universe.
In general these directions can lead to the coherent production of
scalar condensates.
The finite density SUSY breaking has an important impact
on the production of such condensates by defining the
soft potential at early times.
Coherent production of Polonyi type moduli is in general
very dangerous, but a number of ``solutions'' have
been proposed, including enhanced symmetry points, additional
dynamics to give the fields 
a large mass, and late inflation.
Coherent production of standard model fields can give
rise to the baryon asymmetry, and in special cases dark matter
from relic LSPs produced in the decay.
Flat directions are also natural candidates to act as inflatons.
This generally requires an additional hidden sector which breaks
supersymmetry at a very large scale.
Supersymmetry must be restored and the superpotential must
vanish in this sector
at the minimum of the inflaton
potential.

Throughout there have been a couple of hints of possible
cosmological
selection principles.
Symmetries can give technically natural solutions to the Polonyi
problem and the requirements which must be satisfied by the
SUSY breaking sector responsible for inflation.
If this is the case, then our vacuum is near a point of enhanced symmetry.
This might imply there are additional gauge bosons and/or matter
multiplets just above the weak scale (or perhaps some of the
ones we see).
It also might give a cosmological selection
criterion for interesting string vacua.
Unfortunately symmetries are not the unique solutions to the above
requirements.
Even so, we may have glimpses of cosmological selection principles
for the type of vacuum in which we live.

\vspace{.1in}
I would like to thank T. Banks, M. Berkooz, M. Dine, and M. Peskin
for useful discussions and suggestions.
I would also like to M. Dine and L. Randall who collaborated
on some of the work presented here.

\end{document}